\documentclass[%
 reprint,
 amsmath,amssymb,
 aps,
]{revtex4-1}

\usepackage{graphicx}
\usepackage{amsfonts,color}
\usepackage{color}
\usepackage{dcolumn}
\usepackage{bm}

\begin{document}

\preprint{APS/123-QED}

\title{Evading Anderson localization in a one-dimensional conductor with correlated disorder}


\author{Onuttom Narayan}
\affiliation{
Physics Department, University of California, Santa Cruz, CA 95064
}%

\author{Harsh Mathur}
\affiliation{
	        Department of Physics, Case Western Reserve University, Cleveland, OH 44106-7079
}%

\author{Richard Montgomery}
\affiliation{
Mathematics Department, University of California, Santa Cruz, CA 95064
}%



\date{\today}

\begin{abstract}

We show that a one dimensional disordered conductor with correlated disorder has an 
extended state and a Landauer resistance that is non-zero in the limit of infinite
system size in contrast to the predictions of the scaling theory of Anderson localization. 
The delocalization transition is not related to any underlying symmetry of the model such
as particle-hole symmetry. For a wire of finite length the effect manifests as a sharp
transmission resonance that narrows as the length of the wire is increased. Experimental
realizations and applications are discussed including the possibility of constructing
a narrow band light filter. 

\end{abstract}

\maketitle


\section{\label{sec:Intro} Introduction}

In a seminal paper in 1958 Anderson demonstrated that electronic 
states in disordered solids may be localized over a range of
energies \cite{anderson}. Over the next two decades, studies of disordered 
electronic systems culminated in the discovery that in one
and two dimensions electronic states are always localized,
no matter how weak the disorder, while in three dimensions
localized and extended states can exist over different
ranges of energy, separated by a mobility edge \cite{gangone}. 
These findings completely subverted the simple dogma
of band theory, showing that for weakly interacting electrons,
disorder---rather than the
band structure in the clean limit---determined whether
a material is a conductor or insulator at low temperature,
and that at low temperature all weakly interacting materials are insulators
in one and two dimensions \cite{leer}. Moreover the ideas of Anderson 
localization proved relevant to optics, acoustics, cold 
atoms, neural networks, medical imaging and in general 
to any problem of coherent propagation of waves in a random medium \cite{fifty}. 

In the case of one dimension in particular it
has been possible to derive exact results and even rigorous 
proofs of localization for appropriate models \cite{ishii} \cite{gangtwo}. Two 
distinct approaches have been developed to describe 
the universal features of localization. The first 
approach, grounded in random matrix theory, posits
that the distribution of transfer matrices
in one dimension undergoes diffusion in the
space of possible transfer matrices as a function
of the length of the conductor \cite{mello}. This approach, 
which is restricted to one dimension, reveals
that the conductance has a broad log normal
distribution, with very different typical and
mean values, both of which decay exponentially
with the length of the system (a highly non-Ohmic
size dependence). Field theory methods, based
on replicas \cite{stone} or supersymmetry \cite{efetov} for disorder averaging,
likewise describe the universal features of localization
on length scales that are large compared to the microscopic
elastic scattering length, and confirm the picture
described above. Thus localization, particularly 
in one dimension, is now a well-established paradigm.

One known exception to complete localization in
one dimension is systems with particle-hole symmetry \cite{balents}.
In this case it is known that at the symmetric point of
zero energy there is an extended state and hence a 
delocalization transition that separates Anderson insulators
above and below zero energy. More generally the discovery
that quantum systems can be classified into ten symmetry classes \cite{zirnbauer}
based on the absence or presence of particle-hole and time-reversal
symmetries has furnished additional examples of delocalization at zero energy \cite{ilya}.
Another example of an extended state at an isolated energy
is provided by the quantum dimer model \cite{philip}. In this case the
delocalization happens because the individual scatterers
become transparent at a common resonant energy making the
system effectively clean at that energy. Other than that
every attempt to identify extended states (e.g. numerically)
has foundered, strengthening the belief in the inevitability
of localization in one dimension.

Nonetheless in this paper we report the astonishing finding
that for a model of correlated disorder also it is possible to
for the system to undergo delocalization. The extended states
we find do not depend upon the perfect transparency of individual
impurities or on any underlying symmetries of the model but rather
upon local correlations amongst successive scatterers. 
The model consists of symmetric scatterers that are
separated by variable distances. Both the opacity of the
individual scatterers and the spacing between them are 
random variables. However there are correlations amongst
these random variables: the spacing between the successive
scatterers is constrained by the opacities of the scatterers.

The remainder of the paper is organized as follows.
In section II we introduce the model and show that it
can be obtained from an underlying tight binding model
with on-site disorder. In section III 
we demonstrate using a combination of analytic arguments 
and numerical simulations that the
probability of non-zero conductance remains finite no
matter how long the conductor grows contrary to the
localization paradigm. In section IV we offer some 
concluding remarks on possible experimental tests, 
applications and open questions. 

\section{The Model}
\subsection{Individual scatterers}
\label{sec:individual}
To be concrete, we consider a one-dimensional lattice with
non-interacting electrons and nearest neighbor hopping.  The potential
at each site is either zero or is a random number between $-W$ and $W.$ 
The corresponding Schr\"{o}dinger equation is the difference equation
\begin{equation}
    \psi_{n-1} + V_n \psi_n + \psi_{n+1} = E \psi_n.
    \label{eq:schrodinger}
\end{equation}
Here $\psi_n$ is the wavefunction at site $n$ and $V_n$ is the site potential. 
Note that in the absence of disorder ($V_n = 0$ for all $n$) the solutions 
are plane waves $\psi_n = \exp (i k n)$ with energy $E = 2 \cos k$. 
Sites where $V_n \neq 0$ are called scatterers, and we constrain
the system so that two scatterers cannot be adjacent to each other.
Thus the system can be understood as a sequence of scatterers, with
each successive pair of scatterers separated by a lattice segment
where $V=0,$ i.e.  a clean segment.  Our strategy is to find the
$S$-matrix for each scatterer, and combine them appropriately.

\begin{figure}
	\begin{center}
		\includegraphics[width=\columnwidth]{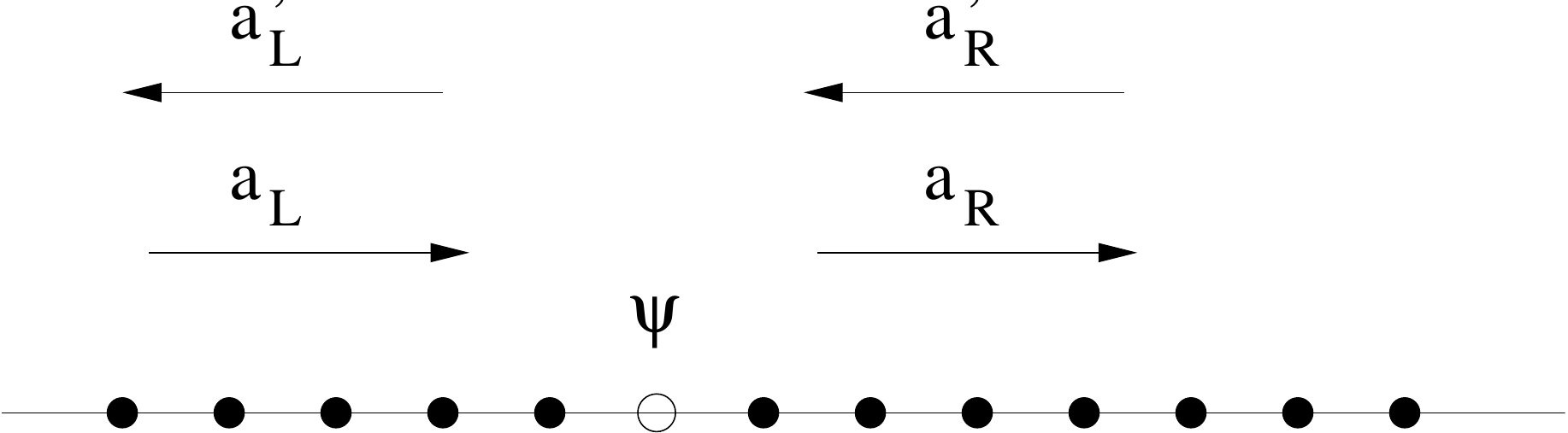} 
		\caption{A scattering site, with incoming and 
		outgoing waves on both sides. The wavefunction at
		the scattering site is $\psi.$ The phases of the
		waves on both sides are chosen so that the amplitude
		would have been $a_{L,R} + a^\prime_{L,R}$ at the
		scattering site if the waves from the left/right
		were to continue uninterrupted through the scattering
		site.}
		\label{fig:singlesite}
	\end{center} 
\end{figure} 
Figure~\ref{fig:singlesite} shows a single scatterer.
We make the ansatz
\begin{eqnarray}
    \psi_n & = & a_L \exp (i k n) + a^\prime_L \exp ( - i k n ) \hspace{2mm} {\rm for} \hspace{2mm} n \leq 0
    \nonumber \\
    & = & a_R \exp (i k n) + a^\prime_R \exp ( - i k n ) \hspace{2mm} {\rm for} \hspace{2mm} n \geq 0
    \label{eq:ansatz}
\end{eqnarray}
Making use of the Schr\"{o}dinger Eq. (\ref{eq:schrodinger}) for $n=0$ we obtain
\begin{eqnarray}
	a_L + a^\prime_L &=& \psi \nonumber\\ a_R + a^\prime_R &=&
	\psi \nonumber\\ (2 \cos k - V) \psi &=& a_L e^{-ik} +
	a^\prime_L e^{i k} + a_R e^{i k} + a^\prime_R e^{-ik}.
\label{eq:ansatzsub}
\end{eqnarray} 
Solving eq (\ref{eq:ansatzsub}) yields
\begin{equation}
	\begin{pmatrix}
		a^\prime_L \\ a_R
	\end{pmatrix} = S \begin{pmatrix}
		a_L \\ a^\prime_R
	\end{pmatrix} 
	\label{Smatrix}
\end{equation} 
Here the $2\times 2$ $S$-matrix connects the outgoing
amplitudes to the incoming amplitudes. Explicitly, we find
\begin{equation}
	S(V, k)  = -\frac{1}{2 i \sin k + V} \begin{pmatrix}
		V  & -2 i \sin k \\ -2 i \sin k & V
	\end{pmatrix}.
	\label{Sone}
\end{equation}

The form of the $S$-matrix for a single scatterer is powerfully constrained by general
principles. Probability conservation imposes unitarity, $S^\dagger S = 1$. 
Parity imposes the additional requirements that
$S_{11} = S_{22}$ and $S_{12} = S_{21}$. 
The most general $2 \times 2$ matrix consistent
with these requirements may be parametrized as
\begin{equation}
	S = e^{i\gamma}
	\begin{pmatrix}
		\cos\theta & i \sin \theta \\
		i \sin \theta & \cos\theta.
	\end{pmatrix}
	\label{Ssymm}
\end{equation}
where $0 \leq \theta \leq \pi/2$ and $0 \leq \gamma < 2 \pi$. 
It follows from Eqs. (\ref{Smatrix}) and (\ref{Ssymm}) that
the transmission coefficient is $\sin^2 \theta$ and the reflection
coefficient is $\cos^2 \theta$. Hence we refer to the parameter
$\theta$ as the opacity of the $S$-matrix. 

Comparison of Eqs. (\ref{Sone}) and (\ref{Ssymm}) shows that 
for the tight binding model analyzed above 
the opacity $\theta$ and the overall phase $\gamma$ for a single scatterer are given by
\begin{eqnarray}
	\exp(i \theta) &=& \pm \frac{V - 2 i \sin k}{\sqrt{V^2 + 4 \sin^2 k}} \nonumber\\
	\exp(i\gamma) &=& \mp \frac{\sqrt{V^2 + 4 \sin^2 k}}{V + 2 i \sin k} = - \exp(i \theta)
	\label{thetagamma}
\end{eqnarray}
where the sign on the right hand side is positive (negative) when
$V$ is positive (negative), i.e. $\cos\theta > 0.$ The fact that
$\theta$ and $\gamma$ are related for this model is not true in general.
This coincidence will play no role in our subsequent analysis. 

\subsection{Combining scatterers}
\begin{figure}
	\begin{center}
		\includegraphics[width=\columnwidth]{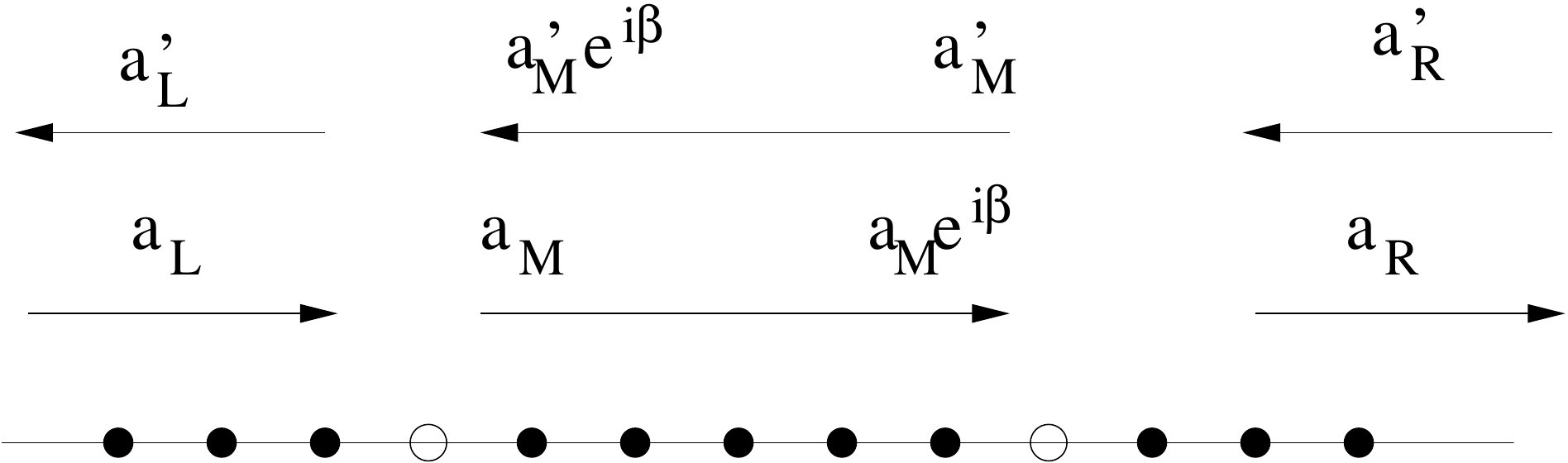} 
		\caption{Two scatterers, with waves propagating to the left 
		and to the right in each region. In the middle region, the left-moving
		wave is incoming for the first scatterer and outgoing for the second 
		scatterer, and vice-versa for the second scatterer. The amplitudes of both
		of these are defined to be zero at the scatterer they start from, and 
		$\exp[i k (L + 1)] = \exp[i\beta]$ at the scatterer they end up at, where 
		$k$ is the wavevector and $L$ is the length of the region between the scatterers;
		in the figure, $L=5.$
		}
		\label{fig:combined}
	\end{center} 
\end{figure} 
The complete lattice can be treated as a sequence of scatterers,
separated by free propagation segments of variable length. Since
the lattice is not left-right symmetric, the $S$-matrix of the
entire system is not as constrained as Eq.(\ref{Ssymm}). Nevertheless,
time reversal invariance requires that $S^* = S^{-1}$ which, together
with the unitarity of $S,$ implies that $S=S^T.$ Therefore we obtain the
parameterization 
\begin{equation}
	S_N  = e^{i\gamma_N} 
	\begin{pmatrix}
		\cos\theta_N e^{i\delta_N} & i \sin\theta_N\\
		i \sin\theta_N & \cos\theta_N e^{-i\delta_N} 
	\end{pmatrix}
\end{equation}
where $S_N$ is the $S$-matrix of a lattice with $N$ scatterers and the parameters
have the domain $0 \leq \theta_N \leq \pi/2$, $0 \leq \delta_N < 2 \pi$ and $0 \leq \gamma_N < 2 \pi$. 

Now suppose that a lattice with $N$ scatterers has an additional impurity attached to its
end with $L$ sites between the $N^{{\rm th}}$ and $(N+1)^{{\rm th}}$ scatterer. 
The amplitudes of the waves in the different regions are as shown in 
Fig \ref{fig:combined}. In the intermediate region the forward
and backward waves have amplitudes $a_M$ and $a^\prime_M$ respectively.
The phases are chosen so that at the site immediately to the right of
the $N^{{\rm th}}$ scatterer the wavefunction is $a_M \exp (i k) + a^\prime_N \exp (i k L )$. 
Hence at the site immediately to the left of the $(N+1)^{{\rm th}}$ 
scatterer the wavefunction is 
$a_M \exp ( i k L) + a^\prime_M \exp ( i k )$. Defining $\exp ( i \beta ) = 
\exp [ i k (L+1) ]$ we have
\begin{equation}
	\begin{pmatrix}
		 a^\prime_M\\ a_R 
	\end{pmatrix}
	= S 
	\begin{pmatrix} 
		a_M e^{i \beta}\\ a^\prime_R
	\end{pmatrix}.
	\label{eq:sone}
\end{equation}
where $S$ is the $S$-matrix of the $(N+1)^{{\rm th}}$ impurity. 
On the other hand, the effect of the $N$ previous scatterers can
be represented as
\begin{equation}
	\begin{pmatrix}
		 a^\prime_L \\ a_M
	\end{pmatrix} 
=S_N
	\begin{pmatrix}
		a_L \\ a^\prime_M e^{i \beta} 
	\end{pmatrix}.
	\label{eq:senn}
\end{equation}
Our objective is to calculate $S_{N+1}$, the $S$-matrix for the combined system,
which is defined by
\begin{equation}
    \left( \begin{array}{c} 
    a^\prime_L \\
    a_R 
    \end{array} \right) = S_{N+1} \left( \begin{array}{c} 
    a_L \\
    a^\prime_R 
    \end{array} \right). 
    \label{eq:snplusone}
\end{equation}
By eliminating the intermediate amplitudes from Eqs. (\ref{eq:sone}) and (\ref{eq:senn}), 
after a lengthy but straightforward calculation, we obtain
\begin{eqnarray}
	\cos\theta_{N+1} e^{i(\gamma_{N+1} + \delta_{N+1})} &=& e^{i(\gamma_N + \delta_N)} \frac{\cos\theta_N  - \cos\theta e^{i \phi}}{1 - 
	\cos\theta \cos\theta_N e^{i\phi}} \nonumber\\
	\cos\theta_{N+1} e^{i(\gamma_{N+1} - \delta_{N+1})} &=& 
	e^{i\gamma}\frac{\cos\theta - \cos\theta_N e^{i \phi}}{1 - \cos\theta \cos\theta_N e^{i\phi}} \nonumber\\
	i \sin\theta_{N+1} e^{i\gamma_{N+1}} &=& - e^{i(\gamma + \gamma_N + \delta_N)/2}\frac{\sin\theta\sin\theta_N e^{i\phi/2}}{1 - \cos\theta \cos\theta_N e^{i\phi}}.\nonumber\\
	\label{monster}
\end{eqnarray}
Here we have defined $\phi = 2\beta + \gamma + \gamma_N - \delta_N.$ 
Note that $\phi$ depends on the phases of the two $S$ matrices being combined
and through $\beta$ also on the distance between the new $(N+1)^{{\rm th}}$ 
scatterer and its predecessor. 

Eq (\ref{monster}) is the main result of this section.
It relates the parameters of the $N+1$ scatterer $S$-marix,
$(\theta_{N+1}, \gamma_{N+1}, \delta_{N+1})$ to $(\theta_N, \gamma_N, \delta_N)$
and $(\theta, \gamma)$ the parameters of the $S$-matrices for the first
$N$ scatterers and for the $(N+1)^{{\rm th}}$ scatterer respectively. 

Although we have couched our discussion in terms of a tight binding 
model it should be obvious that our analysis is much more general. For
example it also applies
to a continuum model in which the scatterers are rectangular
top hat potentials of variable heights and widths separated
by variable distances. The only part of the analysis that would
change is Eq (\ref{thetagamma}) would be replaced expressions
relating the parameters of the $S$ matrix to the barrier 
height and width. 

\section{Delocalization}
\subsection{Analytical results}
In our representation of the $S$-matrix, the transmission coefficient
of the $N$-scatterer lattice is $\sin^2\theta_N.$ By Landauer's formula \cite{harold}
this is the conductance of a system with $N$-scatterers in units of $e^2/h$. 
It follows from the third equality in Eq. (\ref{monster}) 
that the transmission coefficient evolves according to 
\begin{equation}
	\sin^2 \theta_{N+1} = \frac{\sin^2 \theta \sin^2 \theta_N} {1 + \cos^2 \theta \cos^2 \theta_N - 2 \cos\theta \cos\theta_N \cos\phi}.
	\label{iteration}
\end{equation}
Similar relations can be written down that give the phases $\gamma_{N+1}$ and $\delta_{N+1}$ in terms
of the parameters of the matrices $S_N$ and $S$ but for the sake of brevity they are omitted. 
We now describe how Eq. (\ref{iteration}) conventionally leads to Anderson localization and
how suitably correlated disorder may evade it. 

If the scatterers are dilute and randomly distributed then $\phi$ can be treated as a uniform random variable. 
Taking the reciprocal of both sides of eq (\ref{iteration}) and averaging over disorder we obtain 
\begin{equation}
	\langle \csc^2 \theta_{N+1}\rangle = \langle \csc^2 \theta\rangle \langle \csc^2 \theta_N\rangle + \langle \cot^2 \theta\rangle \langle \cot^2 \theta_N\rangle.
	\label{eq:andersoniteration}
\end{equation}
Note that the averages factorize because the opacity $\theta$ and the phase $\gamma$ of the
$(N+1)^{{\rm th}}$ scatterer are random variables independent of the scatterers that preceded it.
In context of the tight-binding model the distribution of $(\theta, \gamma)$ is fixed by 
Eq. (\ref{thetagamma}) and the specified uniform distribution of $V$ over the interval between $-W$ and $W$;
however our conclusions are not limited to this specific choice of distribution. 
The only assumption we have to make is that the probability of extremely opaque scatterers is
small; more precisely that the probability of small opacity $\theta$ goes to zero sufficiently fast
that $\langle \cot^2 \theta \rangle$ is finite which is certainly the case for our tight binding
model or any other reasonable model we might consider. 
Making use of trigonometric identities we may rewrite Eq. (\ref{eq:andersoniteration}) as
\begin{equation}
	\langle\csc^2 \theta_{N+1}\rangle = [1 + 2 \langle \cot^2 \theta\rangle ] \langle \csc^2 \theta_N\rangle -
	\langle \cot^2 \theta\rangle.
\end{equation}
By iterating this relation it is easy to see that $\langle \csc^2 \theta_N \rangle$, which has 
the interpretation of the mean resistance of the sample, grows exponentially with the system size $N$.
This is the essence of Anderson localization. Note that our analysis only shows that the mean resistance grows
exponentially which is not the same as proving that the mean Landauer conductance $\langle \sin^2 \theta_N \rangle$
decays exponentially but all of this is well established lore and is not the focus of our paper (for a 
calculation of the full distribution of the resistance for this model see for example ref \cite{pendry}).

Now let us look for a qualitatively different fixed point for the evolution 
Eq. (\ref{iteration}). To this end at first we assume that all the scatterers
are identical and have the same opacity $\theta$. 
We also assume that the phase $\phi$ can be held constant
for each successive scatterer that is added to the system. With these assumptions 
Eq. (\ref{iteration}) is a simple deterministic map for $\theta_N$ with a fixed point 
$\theta^\ast$ given by
\begin{equation}
	1 = \frac{\sin^2 \theta}{1 + \cos^2 \theta \cos^2 \theta_* - 2 \cos\theta \cos\theta_* \cos\phi}
\end{equation}
As long as $\cos\phi /\cos\theta > 1,$ or equivalently $- \theta < \phi < \theta$, 
this equation has a solution.
The  condition for $\theta_*$ to be a stable  fixed point  is  $-1 < d\sin^2 \theta_{N+1}/d\sin^2 \theta_N
< 1$ at $\theta_N = \theta_{N+1} = \theta_*$ and can be verified to be
always satisfied.

Now let us return to the disordered problem. In this case the 
opacity $\theta$ is drawn from a distribution each time Eq. (\ref{iteration}) 
is evolved. To try to retain the non-trivial solution for the deterministic
case found above we constrain $\phi$ to be a random variable drawn from a 
distribution that satisfies the solvability condition
$-\theta < \phi < \theta$ noted above. Note that if we imagine building the
system one scatterer at a time what we are effectively saying is that the position
of the $(N+1)^{{\rm th}}$ scatterer is constrained to lie within a certain interval 
that is determined by the $S$-matrix of the preceding $N$ scatterers. 
However within that range we can place the $(N+1)^{{\rm th}}$ scatterer at random.
Hence the system we are considering is random but with correlated disorder. 
We now show that this random system evades Anderson localization. 

To this end we again take the reciprocal of both sides of Eq. (\ref{iteration}) 
and average over disorder to obtain
\begin{eqnarray}
	\langle \csc^2 \theta_{N+1}\rangle &=& \langle \csc^2 \theta_N\rangle \langle \csc^2 \theta\rangle 
	+ \langle \cot^2 \theta_N\rangle \langle \cot^2 \theta\rangle \nonumber\\
	&&- 2 \langle \cot\theta_N \csc\theta_N \rangle \langle \cot\theta\csc\theta \cos\phi\rangle.
\end{eqnarray}
We have used the fact that $\theta$ is independent of $\theta_N$ 
and $\phi$ is correlated with $\theta$, not with $\theta_N$. 
For simplicity let us assume that $\phi$ is uniformly distributed over
the interval $-\theta$ to $\theta$. Performing the average of $\phi$ we then obtain 
\begin{eqnarray}
	\langle \csc^2 \theta_{N+1}\rangle &=& \langle \csc^2 \theta_N\rangle \langle \csc^2 \theta\rangle 
	+ \langle \cot^2 \theta_N\rangle \langle \cot^2 \theta\rangle \nonumber\\
	&&- 2 \langle \cot\theta_N \csc\theta_N \rangle \langle \cot\theta/\theta\rangle.
\label{eq:nonanderson}
\end{eqnarray}
With some rearrangement Eq. (\ref{eq:nonanderson}) can be brought to the form
\begin{eqnarray}
    \langle \csc^2 \theta_{N+1} \rangle - \langle \csc^2 \theta_N \rangle & = & 
    - 2 B \langle \csc^2 \theta_N \rangle 
    \nonumber \\
    & + & \left[ \langle \frac{\cot \theta}{\theta} \rangle \langle R (\theta_N) \rangle -
    \langle \cot^2 \theta \rangle \right]. \nonumber \\
    \label{eq:andersontoo}
\end{eqnarray}
Here $B$ is given by
\begin{equation}
    B = \langle \frac{\cot \theta}{\theta} - \cot^2 \theta \rangle = 
    \langle \cot^2 \theta \left( \frac{\tan \theta}{\theta} - 1 \right) \rangle 
    \label{eq:bee}
\end{equation}
and is evidently a finite positive constant 
since $\tan \theta \geq \theta$ over the interval from
zero to $\pi/2$. The specific value of $B$ will depend on the distribution 
chosen for the opacity $\theta$. 
$R(\theta_N) = 2 ( 1 - \cos \theta_N )/ \sin^2 \theta_N$ is a monotonic
decreasing function that goes from 1 to zero as $\theta_N$ goes from zero to $\pi/2$. 
Hence the final term in eq (\ref{eq:andersontoo}) in square brackets is finite
and lies in the range between $- \langle \cot^2 \theta \rangle$ and $B$. 
These observations and the form of Eq. (\ref{eq:andersontoo}) preclude
Anderson localization for this disordered conductor. For a localized conductor
the average resistance should grow monotonically and without bound. 
However if $\langle \csc^2 \theta_N \rangle$ 
gets sufficiently large then the right hand side of Eq. (\ref{eq:andersontoo}) becomes
negative contradicting the assumption that $\langle \csc^2 \theta_N \rangle$
is growing monotonically without bound. Rather if the mean resistance is growing
monotonically Eq. (\ref{eq:andersontoo}) shows that it must saturate to a value
less than unity (in units of $h/e^2$). Even if we relax the assumption of monotonic
behavior Eq. (\ref{eq:andersontoo}) shows that the resistance is bounded which is
incompatible with Anderson localization. Moreover the finiteness of $\langle \csc^2 \theta_N \rangle$
shows that the distribution $P(\theta_N)$ must vanish as $\theta_N \rightarrow 0$. 

We should draw attention to the fact that in order to obtain delocalization we had to
build up our disordered conductor by choosing the phase $\phi$ for each successive
scatterer to lie in an appropriate range of positions. The appropriate range depends
on the energy parameter $k$ so the question arises whether the conductor will remain
delocalized if the energy is varied. While one might hope that the conditions on the
phase might be met for a range of energies in fact our numerical simulations below show that
this is not the case. For a finite system therefore we will get transmission over a narrow
range of energies and exponential localization away from the delocalization energy; 
the transmission resonance will narrow as the system grows.

\subsection{Numerical results}
\begin{figure}
	\begin{center}
		\includegraphics[width=\columnwidth]{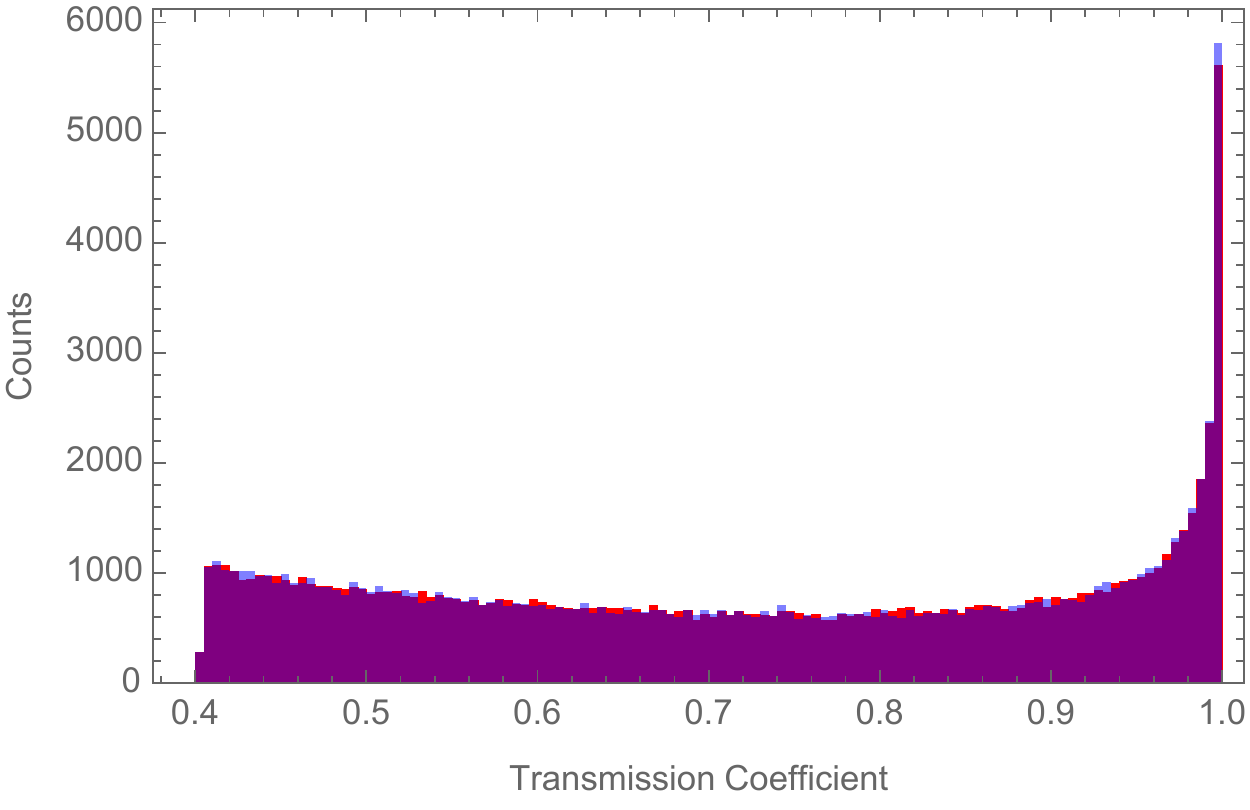}
		\caption{(Color online) Histogram of $\sin^2
		\theta_N.$ Each scatterer has the potential $V$
		chosen at random, uniformly over the interval $[-0.3,
		0.3].$ The angle $\theta$ associated with a scatterer
		is $\cos^{-1} (|V|/\sqrt{V^2 + 4 \sin^2 k}),$ where
		we have chosen the energy $2\cos k $ to be 1.2. The
		phase $\phi$ between each scatterer and its predecessor
		is chosen to be a uniform random variable over the
		interval $0 < \phi < \theta.$ The histogram is
		plotted for $10^5$ random lattices with $N=100$
		scatterers (blue) and $N=10,000$ scatterers (red).
		No significant difference is seen between the two.}
		\label{fig:oneenergy}
	\end{center}
\end{figure}
Figure~\ref{fig:oneenergy} shows numerical results for the transmission
coefficient $\sin^2\theta_N.$ Histograms for $N=100$ and $N=10000$
show no significant difference, indicating that the $N\rightarrow\infty$
limit has been reached. Each scatterer has a $S$-matrix of the form
of Eq.(\ref{Sone}) with $V$ chosen uniformly over the interval $=0.3
< V < 0.3$ and $2 \sin k = 1.6.$ As discussed at the end of
Section~\ref{sec:individual}, the angle $\theta$ for each scatterer
is chosen to be in the first or fourth quadrant, and in
Eq.(\ref{iteration}), it can be chosen to be in the first quadrant
without loss of generality. Since the allowed range of $V$ is small,
all the scatterers are weak, and $\theta$ lies within a small
interval near $\pi/2.$ The angle $\phi$ is chosen randomly, as
described earlier. Under these conditions, one can show analytically
that the distribution for $\sin^2\theta_N$ has a lower cutoff which
is greater than zero, as seen in the figure. 

\begin{figure}[htb]
	\begin{center}
		\includegraphics[width=\columnwidth]{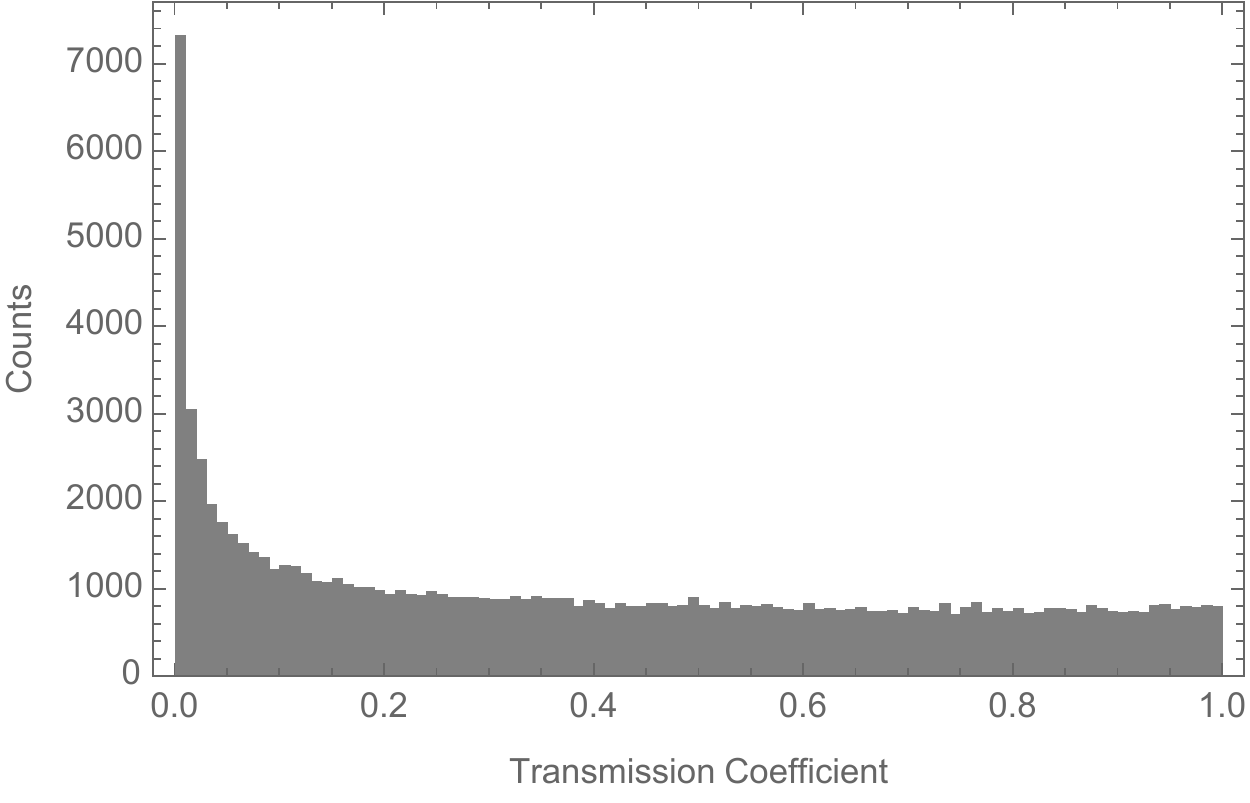}
		\caption{Histogram of $\sin^2 \theta_N.$ The angle
		$\theta$ for each scatterer is chosen to be a uniform
		random variable in the first quadrant, and the angle
		$\phi$ is a uniform random variable between 0 and
		$\theta.$ The histogram is constructed from $10^5$
		lattices with $N=1000$; increasing $N$ does not
		change this significantly. Although there is a peak
		in the histogram at $\sin^2 \theta_N = 0,$ possibly
		a divergence, there is a long tail to the distribution,
		and $\langle \sin^2\theta_\infty\rangle$ is non-zero.
		} \label{fig:uniform}
	\end{center}
\end{figure}
On the other hand, if we consider $\theta$ to be a uniform random
variable in the interval $(0, \pi/2]$ (with $0 < \phi < \theta,$),
$\langle \cot^2\theta\rangle$ diverges and the proof that $\langle
\cot^2 \theta_{N\rightarrow\infty}\rangle$ is finite is not valid.
Nevertheless, as seen in Figure~\ref{fig:uniform}, $\langle \sin^2
\theta_{N\rightarrow\infty}\rangle$ is finite; the peak at the
origin of the distribution is accompanied by a broad flat tail.

From the arguments above, one might think that a random lattice
that is designed to have a $O(1)$ transmission coefficient at a
certain energy should have an $O(1)$ transmission coefficient (i.e.
delocalized states) over a {\it band\/} of energy, since the
$S$-matrix for each scatterer as well as the phase introduced by a
path length $L$ evolve continuously as a function of the wavevector
$k.$ However, this is not the case. The path-dependent phase $\phi$
associated with the interval between the $N^{{\rm th}}$ and $(N+1)^{{\rm th}}$
scatterers was defined to be $\phi = 2 k (L + 1) + \gamma + \gamma_N
- \delta_N.$ Once the length $L$ of the interval has been optimized
for some $k,$ to ensure that $0 < \phi < \theta,$ a small change
in $k$ could change $\gamma_N - \delta_N$ by a large amount if $N$
is large, so that the condition $0 < \phi < \theta$ would no longer
be satisfied. Numerical simulations reveal this to be the case:
although the prescription above allows one to obtain $O(1)$
transmission at any chosen energy, the transmission coefficient
drops off as one moveas away from this energy, and decays as a
function of $N$ for any energy other than the energy for which the
structure is designed.

\section{Conclusion}

We have shown that a one dimensional conductor with correlated 
disorder has an extended state that is unrelated to any symmetry
of the problem and exists despite the fact that the individual scatterers
are not effectively transparent as in previous examples of one dimensional delocalization. 
Conceptually the delocalized structure is constructed scatterer by scatterer
so it is natural to expect that it can be most readily realized 
experimentally as a stacking of films much like a one dimensional
photonic crystal \cite{photonic}. A possible application of such a structure is as
an extremely narrow band filter for light. In contrast to a photonic
crystal the structure does not have to be engineered with precision;
the randomness is in fact essential to the operation of the filter.
Cold atoms are another experimental arena for localization studies
wherein correlated disorder may be realizable \cite{cold}. An interesting analog
of Anderson localization is provided by the phenomenon of 
dynamical localization in kicked quantum rotors \cite{prange}. Whether 
the ideas discussed in this paper can be exported to that context
or generalized to two and higher dimensions are interesting open
questions.


\bibliography{rmt_spectrum}

\end{document}